\documentclass[a4paper,11pt,titlepage,nofootinbib]{revtex4-2}
\pdfoutput=1
\usepackage[T1]{fontenc}
\usepackage{natbib}
\usepackage{soul,xcolor}
\usepackage{diagbox}
\usepackage{mathrsfs,bm}
\usepackage{comment}
\usepackage{multirow,booktabs,graphicx}
\usepackage{amsmath,amssymb}
\usepackage[bookmarks=false,
breaklinks=false,pdfborder={0 0 1},colorlinks=false]
{hyperref}
\hypersetup{
	colorlinks,linkcolor=blue,citecolor=blue,urlcolor=blue}

\def\ba{\begin{eqnarray}}
\def\ea{\end{eqnarray}}
\def\d{{\rm d}}

\newcommand{\beq}{\begin{equation}}
\newcommand{\eeq}{\end{equation}}
\newcommand{\bal}{\begin{aligned}}
\newcommand{\eal}{\end{aligned}}

\begin{document} 

\title{\boldmath Cosmology in warped massive gravity}

\author{Sebastian Garcia-Saenz$^{a}$}
\email{sgarciasaenz@sustech.edu.cn}

\author{Yuxiang Wei$^{a,b}$}
\email{12232895@mail.sustech.edu.cn}

\author{Xue Zhou$^{a}$}
\email{12332935@mail.sustech.edu.cn}

\affiliation{
$^{a}$Department of Physics, Southern University of Science and Technology, \\
Shenzhen 518055, China
}

\affiliation{
$^{b}$School of Science and Engineering, The Chinese University of Hong Kong,\looseness=-1 \\
Shenzhen 518172, China
}

\begin{abstract}
We study the cosmological dynamics and predictions in the theory of warped massive gravity. This set-up postulates a five-dimensional ghost-free massive graviton with a brane-localized four-dimensional massive gravity potential, and has the virtue of raising the strong-coupling scale of the 4D theory. We identify two classes of models that lead to decoupled equations for the scale factor on the brane: one characterized by a particular choice of boundary conditions for the St\"uckelberg fields and one characterized by a special tuning between the coefficients of the 5D and 4D potentials. In the first case, we find interesting solutions including a cosmological bounce without the need of exotic matter. The second case leads to a modified Friedmann equation, and comparison with data shows the potential of the model to alleviate the Hubble tension.
\end{abstract}

\maketitle
\flushbottom

\section{Introduction}
\label{sec:intro}

Tensions in cosmological data have rekindled interest in infrared modifications of general relativity \cite{Abdalla:2022yfr,CosmoVerse:2025txj}. One natural way to achieve this is by postulating a small but non-zero mass for the graviton, which at least conceptually is very minimalistic, not requiring the inclusion of new particles but instead positing the breaking of diffeomorphism invariance. In keeping with this principle, one would like the breaking to be such that no extra degrees of freedom are introduced beyond those of a massive spin-2 particle, a requirement that turns out to make the problem highly non-trivial \cite{Boulware:1972yco}. Nevertheless, a solution does exist, as given by the so-called de Rham-Gabadadze-Tolley (dRGT) class of massive gravity theories \cite{deRham:2010ik,deRham:2010kj}; see \cite{Hinterbichler:2011tt,deRham:2014zqa} for reviews.

A notorious issue with massive gravity theories is the very low strong-coupling scale of self-interactions, resulting in a limited domain of applicability \cite{Arkani-Hamed:2002bjr}. The root of the problem lies in the helicity-0 mode of the graviton, which on flat spacetime comes in the Lagrangian without standard kinetic term; this means that, upon going to a canonical field basis, the couplings of this mode will be enhanced by inverse powers of the graviton mass. One resolution is to formulate the theory on a curved spacetime, such as anti-de Sitter (AdS) space, which does generate a kinetic term for the helicity-0 mode and indeed results in a higher strong-coupling scale \cite{Kogan:2000uy,Porrati:2000cp,Karch:2001jb} (see also \cite{deRham:2016plk,Bonifacio:2016blz,DeRham:2018axr}). However, AdS space does not seem to provide a good model of our universe.

An interesting alternative has been proposed by Gabadadze \cite{Gabadadze:2017jom}. The AdS space of massive gravity may be higher-dimensional, while the effective four-dimensional theory may resemble massive gravity on a flat background. He has given a concrete realization of this idea: 5D dRGT massive gravity with AdS fiducial metric along with gravitational and dRGT terms localized on a 4D brane. This is the so-called theory of warped massive gravity (WMG), which may be regarded as a massive generalization of the Dvali-Gabadadze-Porrati (DGP) model \cite{Dvali:2000hr}.

In this paper we wish to study the implications of WMG for the cosmology on the brane. A virtue of the DGP model in the cosmological context is that one can derive a decoupled evolution equation for the scale factor on the brane, in the form of a modified Friedmann equation \cite{Binetruy:1999hy,Binetruy:1999ut,Shiromizu:1999wj,Deffayet:2000uy,Deffayet:2001pu}.\footnote{See \cite{deRham:2010tw,deRham:2011by,Comelli:2011zm,DAmico:2011eto,DeFelice:2013bxa,Gong:2012yv,Hinterbichler:2013dv,Volkov:2013roa} for seminal papers on the cosmology of massive gravity.} In other words, one can study the cosmology on the brane without solving the bulk equations of motion, a very significant simplification given that the latter are non-linear PDEs. One of our goals is to see under what conditions the same holds in WMG. We find that the presence of St\"uckelberg fields in general forbids such simplification. However, we identify two particular classes of models which evade the issue, leading to self-consistent evolution equations for the 4D metric scale factor together with the standard conservation equation for brane-localized matter.

In the first model, which we refer to as `Neumann', one chooses trivial Neumann boundary conditions on the brane for the St\"uckelberg functions. This choice is shown to be consistent with the constraint equations of dRGT massive gravity and to produce an independent evolution equation for the scale factor on the brane upon making use of appropriate junction conditions on the bulk metric. Interestingly, however, this equation is not of the Friedmann-type, rather it is a modified Raychaudhuri equation, which we argue cannot be integrated to produce a first-order equation. Our study of solutions, based on analytical and numerical calculations, reveals interesting results, including cosmological bounces without the need of exotic matter as well as Big Crunch models with ordinary matter and \textit{negative} spatial curvature.

The second class of models, which we call `special', is defined by a particular relation between the 5D and the 4D coefficients of the dRGT potentials. This set-up also results in a self-consistent solution of the dRGT constraints and leads to a simplified version of the Raychaudhuri equation which, in this case, does admit a first integral, i.e.\ a modified Friedmann equation, in analogy with the DGP model. The WMG case however introduces modifications, allowing for some novel, if peculiar, solutions. More generically, the model resembles general relativity with a cosmological constant at late times, but introduces modifications at early times. We perform a first analysis of data comparison in this set-up, making use of up-to-date cosmic microwave background and supernova datasets. Interestingly, our results show the potential of the model to alleviate the so-called Hubble tension.

The paper is organized into two main sections, Sec.\ \ref{sec:wmg} on the theoretical analysis and Sec.\ \ref{sec:data analysis} on the data analysis. These may be read independently should the reader be interested on one of these aspects only. Sec.\ \ref{sec:wmg} consists of a brief review of WMG (Sec.\ \ref{sec:wmg review}) followed by our analysis of the  equations of motion and constraints in the context of cosmology (Sec.\ \ref{sec:flrw} and \ref{sec:brane cosmology}). The Neumann and special models are explored in Sec.\ \ref{subsec:neumann model} and Sec.\ \ref{subsec:special model}, respectively. We close with an outlook and a summary of our scope and assumptions in Sec.\ \ref{sec:conclusion}.

%====================================

\newpage

\section{Cosmological spacetimes in warped massive gravity} \label{sec:wmg}

\subsection{Warped massive gravity} \label{sec:wmg review}

In the theory of WMG one considers a 4-dimensional brane in a 5-dimensional bulk spacetime. The action includes ghost-free massive gravity terms, one in the bulk and one localized on the brane:
\beq\bal \label{eq:wmg full action}
S&=\frac{M_5^3}{2}\int \d^5X\sqrt{-g_{(5)}}\left(R_{(5)}-2\Lambda_5\right)+M_5^3m_5^2\int \d^5X\sqrt{-g_{(5)}}\,U_{(5)}(S_{(5)})\\
&\quad+\frac{M_4^2}{2}\int \d^4x\sqrt{-g}\,R+M_4^2m_4^2\int \d^4x\sqrt{-g}\,U(S_{(4)})+S_{\mathrm{brane}} \,.
\eal\eeq
Here $M_4$ and $M_5$ are the 4- and 5-dimensional Planck scales, and $m_4$ and $m_5$ are the 4- and 5-dimensional graviton masses, respectively. The bulk cosmological constant, $\Lambda_5$, is assumed negative, while for simplicity we assume the absence of a cosmological constant in the brane action.\footnote{Note that in Ref.\ \cite{Gabadadze:2017jom} a (positive) brane cosmological constant is needed in order to obtain the desired AdS solution, similarly to the Randall-Sundrum set-up \cite{Randall:1999ee,Randall:1999vf}. Here we are interested instead in more general FLRW solutions. In any case, there is no loss of generality since a 4D cosmological constant may always be included in the matter action $S_{\rm brane}$.} In addition, we consider standard matter fields localized on the brane, with action $S_{\rm brane}$.

The action \eqref{eq:wmg full action} includes the dRGT potentials \cite{deRham:2010kj,Nieuwenhuizen:2011sq}\footnote{This parametrization ensures the absence of tadpole, the normalization of the graviton mass (i.e.\ $m_5^2$ and $m_4^2$ respectively for the bulk and brane) and the absence of cosmological constant (which, if present, is by convention included in the Einstein-Hilbert term or in the matter action).}
\beq\bal
U_{(5)}(S_{(5)})&=e_2(S_{(5)})+\beta_3e_3(S_{(5)})+\beta_4e_4(S_{(5)})+\beta_5e_5(S_{(5)}) \,,\\
U(S_{(4)})&=e_2(S_{(4)})+\alpha_3e_3(S_{(4)})+\alpha_4e_4(S_{(4)}) \,,
\eal\eeq
where $S_{(5)}\equiv \mathbf{1}_{(5)}-\sqrt{g_{(5)}^{-1}f_{(5)}}$ and $S_{(4)}\equiv \mathbf{1}_{(4)}-\sqrt{g^{-1}f}$ (here $\mathbf{1}_{(n)}$ is the $n$-dimensional identity matrix), and $e_n(M)$ denotes the $n$-th elementary symmetric polynomial of the eigenvalues of the matrix $M$. The dimensionless constants $\alpha_n$ and $\beta_n$ are free parameters of the model. Notice that, by definition, $g_{\mu\nu}$ is the pullback of the bulk dynamical metric $g^{(5)}_{MN}$ (similarly to the DGP set-up) and $f_{\mu\nu}$ is the pullback of the bulk fiducial metric $f^{(5)}_{MN}$.\footnote{In our notations, $M,N,\ldots\in\{0,1,2,3,4\}$ and $\mu,\nu,\ldots\in\{0,1,2,3\}$ denote 5- and 4-dimensional spacetime indices, respectively, while $i,j,\ldots\in\{1,2,3\}$ denote 3-dimensional spatial indices.}

Recall that the fiducial metric is non-dynamical, and is an input in the definition of the theory. In WMG it is taken to be an AdS metric with a flat slicing (i.e.\ in Poincar\'e coordinates) so that the induced fiducial metric is simply Minkowski,
\beq \label{eq:fiducial metric}
\d s^2_{f(5)}=\frac{L^2}{(Y+L)^2}\left(\eta_{\mu\nu}\d X^{\mu}\d X^{\nu}+\d Y^2\right) \,,\qquad \d s^2_{f(4)}=\eta_{\mu\nu}\d X^{\mu}\d X^{\nu} \,,
\eeq
in a gauge where the brane is located at $Y=0$, and $L=\sqrt{-\frac{6}{\Lambda_5}}$ is the AdS radius.

The metric $g^{(5)}_{MN}$ is determined dynamically from the field equations. Variation of the action \eqref{eq:wmg full action} yields
\beq \label{eq:5d field equations}
G^{(5)}_{MN}+\Lambda_5g^{(5)}_{MN}+m_5^2U^{(5)}_{MN}=\kappa {\mathcal{T}}_{MN} \,,
\eeq
where $\kappa\equiv M_5^{-3}$ and
\beq \label{eq:tau tensor def}
{\mathcal{T}}_{MN}=\delta(Y)\sqrt{g/g_{(5)}}\,\tau_{\mu\nu}\delta^{\mu}_M\delta^{\nu}_N \,,\qquad \mbox{where}\quad \tau_{\mu\nu}\equiv T_{\mu\nu}-\mu^{-1}\left(G_{\mu\nu}+m_4^2U_{\mu\nu}\right) \,.
\eeq
Here $\mu\equiv M_4^{-2}$, $G^{(5)}_{MN}$ and $G_{\mu\nu}$ are respectively the 5- and 4-dimensional Einstein tensors, and $T_{\mu\nu}$ is the matter energy-momentum tensor (obtained as usual by varying $S_{\rm brane}$ with respect to $g_{\mu\nu}$.) The tensor $U_{\mu\nu}$ is obtained by differentiating the dRGT potential (see e.g.\ \cite{Hassan:2011vm}),
\beq\bal
\;&U_{\mu\nu}\equiv \frac{2}{\sqrt{-g}}\frac{\partial(\sqrt{-g}\,U(S_{(4)}))}{\partial g^{\mu\nu}}=-\sum_{n=2}^4\alpha_n\left[e_n(S_{(4)})g_{\mu\nu}+Y^{(n)}_{\mu\nu}(S_{(4)})\right] \,, \\
&\mbox{where}\quad Y^{(n)}_{\mu\nu}(S_{(4)})=g_{\mu\lambda}\sum_{m=1}^n(-1)^{m}e_{n-m}(S_{(4)})\left[(S_{(4)})^{m-1}\left((S_{(4)})-\mathbf{1}_{(4)}\right)\right]^{\lambda}_{\phantom{\lambda}\nu} \,,
\eal\eeq
with an analogous expression for $U^{(5)}_{MN}$.

\subsection{FLRW spacetimes} \label{sec:flrw}

We consider now a homogeneous and isotropic (in the 3D spatial directions of the brane) ansatz for the dynamical metric. Actually, as is well known from studies in massive gravity, flat and closed FLRW metrics are inconsistent with the equations of motion in the presence of generic matter sources, assuming a flat fiducial metric \cite{DAmico:2011eto}. This applies also in our set-up upon considering the equations on the brane. This no-go result can be evaded however in the case of an open FLRW metric \cite{Gumrukcuoglu:2011ew}. We therefore focus on the line element
\beq \label{eq:5d flrw metric}
\d s^2_{g(5)}=-n^2(t,y)\d t^2+a^2(t,y)\gamma_{ij}\d x^i\d x^j+b^2(t,y)\d y^2 \,,
\eeq
where
\beq
\gamma_{ij}\d x^i\d x^j=\d\bm{x}^2-\frac{K(\bm{x}\cdot\d\bm{x})^2}{1+K|\bm{x}|^2} \,,\qquad \bm{x}\equiv(x^1,x^2,x^3) \,.
\eeq
The fifth dimension is parametrized by the coordinate $y$ and we choose a gauge where the brane is placed at $y=0$. The induced metric is given by
\beq
\d s^2_{g(4)}=-\d t^2+a^2(t)\gamma_{ij}\d x^i\d x^j \,,
\eeq
upon choosing the time coordinate such that $n(t,0)=1$,\footnote{This completely fixes the gauge associated to time reparametrizations, as we have assumed the absence of an off-diagonal $t-y$ term in the metric.} and we write $a(t)\equiv a(t,0)$. This corresponds to the open FLRW metric expressed in Cartesian-like coordinates, with curvature constant $K>0$.\footnote{Note that both $K$ and $x^M$ are dimensionful in our conventions. The metric functions $n(t,y)$, $a(t,y)$ and $b(t,y)$ are dimensionless.}

The flat 4D fiducial metric admits an open FLRW chart given by
%\SGS{Note: $f$ and $g$ have dimensions of length}
\beq
X^0=f(t,y)\sqrt{1+K|\bm{x}|^2} \,,\qquad X^i=f(t,y)\sqrt{K}x^i \,,\qquad Y=g(t,y) \,.
\eeq
In accordance with our gauge choices, we require $g(t,0)=0$ (which also implies $\dot{g}(t,0)=0$, with a dot denoting derivative with respect to $t$). Evaluating on the brane one obtains, as claimed,
\beq \label{eq:4d fiducial metric open chart}
\d s^2_{f(4)}=-\dot{f}^2(t)\d t^2+Kf^2(t)\gamma_{ij}\d x^i\d x^j \,,
\eeq
with $f(t)\equiv f(t,0)$.

The functions $f(t,y)$ and $g(t,y)$ are arbitrary at this stage, except for the aforementioned boundary condition on $g$. They are however constrained by the equations of motion. The cleanest way to obtain these constraints is by adopting a St\"uckelberg formulation of the theory, and varying the action with respect to the St\"uckelberg fields. For our purposes, it is sufficient to only consider the variation with respect to the function $f(t)$, which indeed plays the role of a St\"uckelberg field in the mini-superspace action. Doing so we find the following equation:
\beq \label{eq:f constraint}
\left[Kf^2(\alpha_3+\alpha_4)-2Kaf(1+2\alpha_3+\alpha_4)+a^2(3+3\alpha_3+\alpha_4)\right]\left(\sqrt{K}-\dot{a}\right)=0 \,.
\eeq
Since $a(t)$ is assumed non-constant, we obtain the desired expression for $f(t)$,
\beq \label{eq:f solution}
f_\pm(t)=\frac{a(t)}{\sqrt{K}}A_\pm \,,\qquad A_\pm\equiv \frac{1+2\alpha_3+\alpha_4\pm\sqrt{1+\alpha_3+\alpha_3^2-\alpha_4}}{\alpha_3+\alpha_4} \,.
\eeq
We observe that there are two inequivalent branches of solutions, provided of course $A_\pm$ is real, as we assume. We will further suppose $A_\pm>0$ so that $f(t)$ is positive. This agrees with the result obtained in 4D massive gravity \cite{Gumrukcuoglu:2011ew}.

As already mentioned, the equations of motion derived in the St\"uckelberg theory are equivalent to the constraint equations inferred from the unitary gauge formulation (see e.g.\ \cite{Hassan:2011vm}). Importantly, on the other hand, one should also check the consistency with the condition of energy-momentum conservation of matter. Indeed, using the 5D Bianchi identity in the divergence of \eqref{eq:5d field equations} we have
\beq \label{eq:div U5}
m_5^2\nabla_MU^{(5)M}{}_N=\kappa\nabla_M\mathcal{T}^M{}_N \,,
\eeq
and
\beq
\nabla_M\mathcal{T}^M{}_N=\frac{\delta(y)}{b}\nabla_{\mu}T^{\mu}{}_\nu\,\delta^{\nu}_N \,,
\eeq
which follows because $\nabla^{\mu}U_{\mu\nu}=0$ \textit{identically} on the solution \eqref{eq:f solution}. In fact, the matrix $U^{\mu}{}_\nu\equiv g^{\mu\rho}U_{\rho\nu}$ is constant and proportional to the identity,
\beq\bal \label{eq:cal A def}
\;&U^{\mu}{}_\nu=-\mathcal{A}\,\delta^{\mu}_{\nu} \,, \\
\:&\mbox{where}\quad \mathcal{A}\equiv \frac{(1+\alpha_3)\left(2+\alpha_3+2\alpha_3^2-3\alpha_4\right)+2\sigma\left(1+\alpha_3+\alpha _3^2-\alpha _4\right)^{3/2}}{(\alpha_3+\alpha_4)^2} \,,
\eal\eeq
and here $\sigma=\pm1$ is the sign defining the two branches in \eqref{eq:f solution}. Thus $\nabla_M\mathcal{T}^M{}_N$ is also identically zero provided the brane-localized matter obeys the standard conservation law. Then \eqref{eq:div U5} reduces to the constraint $\nabla_MU^{(5)M}{}_N=0$ for the 5D massive gravity potential. In our set-up, this corresponds to two independent equations that in principle determine the functions $f(t,y)$ and $g(t,y)$ in the bulk. We will come back to this constraint after discussing the field equations on the brane.

\subsection{Cosmology on the brane} \label{sec:brane cosmology}

\subsubsection{Junction conditions}

We are interested in the equations of motion evaluated on the brane. We follow the standard methodology as exposed for instance in Refs.\ \cite{Binetruy:1999hy,Binetruy:1999ut}. While the metric functions must be continuous across the brane, their derivatives exhibit discontinuities, in particular (here a prime means differentiation with respect to $y$)
\beq
[n'](t)\equiv n'(t,0^+)-n'(t,0^-) \,,\qquad [a'](t)\equiv a'(t,0^+)-a'(t,0^-) \,,
\eeq
are non-zero. We refer to these as jump functions. The limit $y\to0^+$ of the equations of motion may then be expressed in terms of these jump functions, for instance $a'(t,0^+)=\frac{1}{2}[a'](t)$, noting that $a'(t,0^-)=-a'(t,0^+)$ because of the $\mathbb{Z}_2$ symmetry of the bulk about the brane.

The jump functions may be obtained systematically from the Israel junction condition \cite{Israel:1966rt},
\beq \label{eq:israel junction condition}
[K_{\mu\nu}]=-\kappa\left(\tau_{\mu\nu}-\frac{1}{3}\tau^{\rho}{}_{\rho}g_{\mu\nu}\right) \,,
\eeq
where $K_{MN}=\nabla_Mn_N-n_Mn^L\nabla_Ln_N$ is the extrinsic curvature of the brane, with $n^M$ the unit vector normal to the brane. Notice that the Israel condition in massive gravity is the same as in general relativity because the dRGT potentials do not involve derivatives of the metric and are thus continuous across the brane. The right-hand side of \eqref{eq:israel junction condition} is however modified by the presence of the 4D gravitational term, which contributes to the `effective' energy-momentum tensor $\tau_{\mu\nu}$, cf.\ \eqref{eq:tau tensor def}.

For the metric \eqref{eq:5d flrw metric} we have $n^M=\delta^M_y/b$ and
\beq
K^M{}_N=\frac{1}{b}\,\mathrm{diag}\left(\frac{n'}{n},\frac{a'}{a},\frac{a'}{a},\frac{a'}{a},0\right) \,.
\eeq
The FLRW isometries imply that the energy-momentum tensor must have a perfect fluid form: $T^{\mu}{}_{\nu}=\mathrm{diag}\left(-\rho,p,p,p\right)$. Substituting everything into \eqref{eq:israel junction condition} we eventually obtain the following expressions for the jump functions:
\beq\bal \label{eq:jump functions}
\frac{[n']}{b}&=\frac{\kappa}{3}(2\rho+3p)+\frac{\kappa}{\mu}\left(2\frac{\ddot{a}}{a}-H^2+\frac{K}{a^2}\right)+\frac{\kappa m_4^2}{3\mu}\mathcal{A} \,,\\
\frac{[a']}{ab}&=-\frac{\kappa}{3}\rho+\frac{\kappa}{\mu}\left(H^2-\frac{K}{a^2}\right)+\frac{\kappa m_4^2}{3\mu}\mathcal{A} \,,\\
\eal\eeq
where $H(t)\equiv \dot{a}/a$ is the Hubble parameter and note that here all metric functions are evaluated on the brane. We also remind the reader of the definition of the constant $\mathcal{A}$ in \eqref{eq:cal A def}.

\subsubsection{Cosmological evolution equation}

The next step is to substitute the jump functions into the $y\to0$ limit of the fields equations \eqref{eq:5d field equations}. These include, in addition to the scale factor, the jump functions $[n']$ and $[a']$ as well as those of $f'$, $g'$ and $b'$. However, two among the components of $G^{(5)}_{MN}$ are independent of $b$, namely $G^{(5)}_{0y}$ and $G^{(5)}_{yy}$. The former yields an equation which is identically satisfied,\footnote{Explicitly, $U^{(5)}_{0y}=0$ and $G^{(5)}_{0y}\Big|_{y\to0}=\dot{a}\frac{[n']}{b}-\frac{\d}{\d t}\left(\frac{[a']}{b}\right)$, and the latter may be easily seen to yield zero upon substitution of the jump functions \eqref{eq:jump functions} and the condition of energy conservation for matter.} while from the latter we find an equation relating $a(t)$ and the jumps.

On the other hand, it is not possible in general to obtain an evolution equation for the scale factor that is independent of the so-far unspecified jump functions $[f']$ and $[g']$. These are determined by boundary conditions, and here we make the simplest choice $[f']=0$. Interestingly, this choice is also sufficient to eliminate any dependence on $[g']$ from the $(yy)$ component of the field equation, with the following result:
\beq \label{eq:modified raych}
\frac{\ddot{a}}{a}+H^2-\frac{K}{a^2}-\frac{1}{4}\left[\left(\frac{[a']}{ab}\right)^2+\frac{[a'][n']}{ab^2}\right]+m_5^2\left[\mathcal{B}_1\frac{\dot{a}}{\sqrt{K}}+\mathcal{B}_2\right]-\frac{\Lambda_5}{3}=0 \,,
\eeq
where
\beq\bal
\mathcal{B}_1&=\frac{A_\pm}{3}\big[-(4+6\beta_3+4\beta_4+\beta_5)+3A_\pm(1+3\beta_3+3\beta_4+\beta_5) \\
&\quad -3A_\pm^2(\beta_3+2\beta_4+\beta_5)+A_\pm^3(\beta_4+\beta_5)\big] \,,\\
\mathcal{B}_2&=-\frac{1}{3}\big[-(10+10\beta_3+5\beta_4+\beta_5)+3A_\pm(4+6\beta_3+4\beta_4+\beta_5) \\
&\quad -3A_\pm^2(1+3\beta_3+3\beta_4+\beta_5)+A_\pm^3(\beta_3+2\beta_4+\beta_5)\big] \,.\\
\eal\eeq
Substitution of the jumps \eqref{eq:jump functions} into \eqref{eq:modified raych} finally yields a second order equation for the scale factor $a(t)$. This is our main result concerning the cosmology on the brane. The equation must be of course supplemented by the matter field equations. In the case of a perfect fluid, this corresponds to the conservation of energy relation, $\dot{\rho}+3H(\rho+p)=0$, together with the fluid equation of state.

Eq.\ \eqref{eq:modified raych} may be interpreted as a `modified Raychaudhuri equation', since indeed it is easy to see that it reduces to the Raychaudhuri equation of 4D GR cosmology in the limit $\kappa\to\infty$ and $m_4\to0$. If instead one keeps $\kappa$ finite and set $m_4=0$ and $m_5=0$ then \eqref{eq:modified raych} reproduces the known result obtained in the DGP model \cite{Binetruy:1999hy,Deffayet:2000uy}. In the massive case, we observe that the equation has additional constant terms proportional to $m_4^2$ and $m_5^2$, giving rise to an `effective' cosmological constant,
\beq \label{eq:lambda eff}
\Lambda_{\rm eff}=\Lambda_5-3m_5^2\mathcal{B}_2+\frac{1}{6}\left(\frac{\kappa m_4^2}{\mu}\mathcal{A}\right)^2 \,,
\eeq
consistent with the expectation that massive gravity should accommodate self-accelerating cosmologies \cite{Gumrukcuoglu:2011ew} (and recall that for simplicity we assume the absence of a 4D cosmological constant, contained in principle in the brane energy density $\rho$). We note however that this $\Lambda_{\rm eff}$ need not agree with the effective cosmological constant that appears in the corresponding modified Friedmann equation, as we shall make explicit in Sec.\ \ref{subsec:special model}.

Remarkably, on the other hand, 5D massive gravity introduces a novel term in the equation, $\propto m_5^2\frac{\dot{a}}{\sqrt{K}}$. There are two interesting aspect about this term: \textit{(i) it is singular in the flat FLRW limit, $K\to0$, and (ii) it prevents the equation from admitting a ``simple'' first integral, i.e.\ this theory does not have a modified Friedmann equation when $\mathcal{B}_1\neq0$} (the case $\mathcal{B}_1=0$ will be discussed in Sec.\ \ref{subsec:special model}).

Regarding property (i) we remark that, although the fiducial metric is singular in the limit $K\to0$, cf.\ \eqref{eq:4d fiducial metric open chart}, this is not observable insofar as matter only couples to the dynamical metric, which on the brane is determined by \eqref{eq:modified raych}, and this equation does admit a well-defined flat FLRW limit when $\mathcal{B}_1=0$. In this sense, the novel term $\propto m_5^2\frac{\dot{a}}{\sqrt{K}}$ drastically changes this conclusion.

To explain property (ii), we note that Eq.\ \eqref{eq:modified raych} must admit a first integral as it can be recast as an autonomous two-dimensional differential system \cite{McCauley_1997}. However, such integral is only guaranteed to exist locally, and even then one may not be able to find $H(a)$, i.e.\ a Friedmann-type equation, expressed in terms of elementary functions. We remark that the problem of classifying two-dimensional systems admitting a globally defined first integral has not been fully solved, even in the polynomial case \cite{LLIBRE2004437} (and \eqref{eq:modified raych} is not polynomial). In Appendix \ref{app:first integral} we show that, indeed, Eq.\ \eqref{eq:modified raych} in general does not admit a first integral allowing one to express $H(a)$ in closed form.\footnote{Recall that in 4D massive gravity, as in GR, the Friedmann equation is equivalent to the Hamiltonian constraint. This may seem at odds with our claim, since WMG by construction certainly does possess a Hamiltonian constraint~\cite{Hassan:2011hr,Hassan:2011tf,Hassan:2011ea}. The point is that this constraint is now a bulk function, which is not guaranteed to yield a related constraint on the brane that is decoupled from the bulk dynamics.}

\subsubsection{Constraint equations}

It remains to consider the massive gravity constraint $\nabla_MU^{(5)M}{}_N=0$. As mentioned, this should be thought of as equations of motion for the St\"uckelberg functions $f$ and $g$. Once again here we are concerned with the equations on the brane only. With the condition $[f']=0$ we eventually find
\beq\bal \label{eq:brane constraints}
0=\nabla_MU^{(5)M}{}_0\Big|_{y\to0}&=\frac{A_{\pm}\dot{a}}{\sqrt{K}ab}\bigg[3\mathcal{C}_1\left(\sqrt{K}-\dot{a}\right)b-(\mathcal{C}_1+\mathcal{C}_2A_{\pm})a\dot{b}+\frac{3}{2}\mathcal{C}_2\left(\sqrt{K}-\dot{a}\right)[g']\bigg] \,,\\
0=\nabla_MU^{(5)M}{}_y\Big|_{y\to0}&=-\frac{[g']}{2}\bigg[\frac{3}{2}\left(\mathcal{C}_1+\mathcal{C}_2A_{\pm}\frac{\dot{a}}{\sqrt{K}}\right)\frac{[a']}{ab}+\frac{1}{2}(\mathcal{C}_1+\mathcal{C}_2A_{\pm})\frac{[n']}{b} \\
&\quad +\frac{A_{\pm}}{L}\left(3\mathcal{C}_1+(\mathcal{C}_1+4\mathcal{C}_2A_{\pm})\frac{\dot{a}}{\sqrt{K}}\right)\bigg] \,,
\eal\eeq
where
\beq\bal \label{eq:C coeffs}
\mathcal{C}_1&=(4+6\beta_3+4\beta_4+\beta_5)-2A_\pm(1+3\beta_3+3\beta_4+\beta_5)+A_\pm^2(\beta_3+2\beta_4+\beta_5) \,,\\
\mathcal{C}_2&=-(1+3\beta_3+3\beta_4+\beta_5)+2A_\pm(\beta_3+2\beta_4+\beta_5)-A_\pm^2(\beta_4+\beta_5) \,.
\eal\eeq

If $[g']\neq0$, we observe that the second of these would result in an evolution equation for the scale factor. However, it is easy to see that this equation is inconsistent with \eqref{eq:modified raych}. We are therefore led to two possibilities: either $[g']=0$ or $\mathcal{C}_1=\mathcal{C}_2=0$, so that the second constraint is identically satisfied.

We refer to these two cases as `Neumann' and `special' models, respectively. The reason behind these names is that the former case is defined by the Neumann boundary conditions $[f']=[g']=0$ on the St\"uckelberg functions, while the latter is defined by a special relation between the 5D and 4D massive gravity coefficients (see Eq.\ \eqref{eq:special beta sol} below). We discuss the two models in the next subsections. As for the first constraint in \eqref{eq:brane constraints}, we see that it is also identically satisfied for the special model, while in the Neumann model it results in a constraint equation that determines $b$ in terms of $a$ on the brane,\footnote{The solution is given by $b(t)= \exp\left[\frac{3\mathcal{C}_1}{\mathcal{C}_1+\mathcal{C}_2A_{\pm}}\int^t\frac{\sqrt{K}-\dot{a}(t')}{a(t')}dt'\right]$.} which must be taken into account if one were to solve the bulk equations for all variables.

\subsection{Neumann model} \label{subsec:neumann model}

Eq.\ \eqref{eq:modified raych} is highly non-linear and we leave an in-depth analytical study to future work. Here we content ourselves with some general remarks and a brief numerical study.

We are particularly interested in the novel term $\propto m_5^2\frac{\dot{a}}{\sqrt{K}}$ highlighted previously. We attempt to isolate the effects of this term by considering the limit $K\to0$. In practice we do so by taking all parameters to be of order one in units of the 4D Planck scale, with the exception of $K\mu\ll1$ (recall that $\mu\equiv M_4^{-2}$). For simplicity we consider a pure dust model, i.e.\ $\frac{\mu}{3H_0^2}\rho=\frac{\Omega_{\rm m}}{a^3}$,\footnote{Throughout the paper, density parameters $\Omega_i$ correspond to the present-day values.} although we have checked that the qualitative features of the model are unchanged upon including radiation. We also set $\Lambda_{\rm eff}=0$ as here we are primarily interested in the early-time behavior.

\begin{figure}
\centering
\includegraphics[width=0.65\textheight]{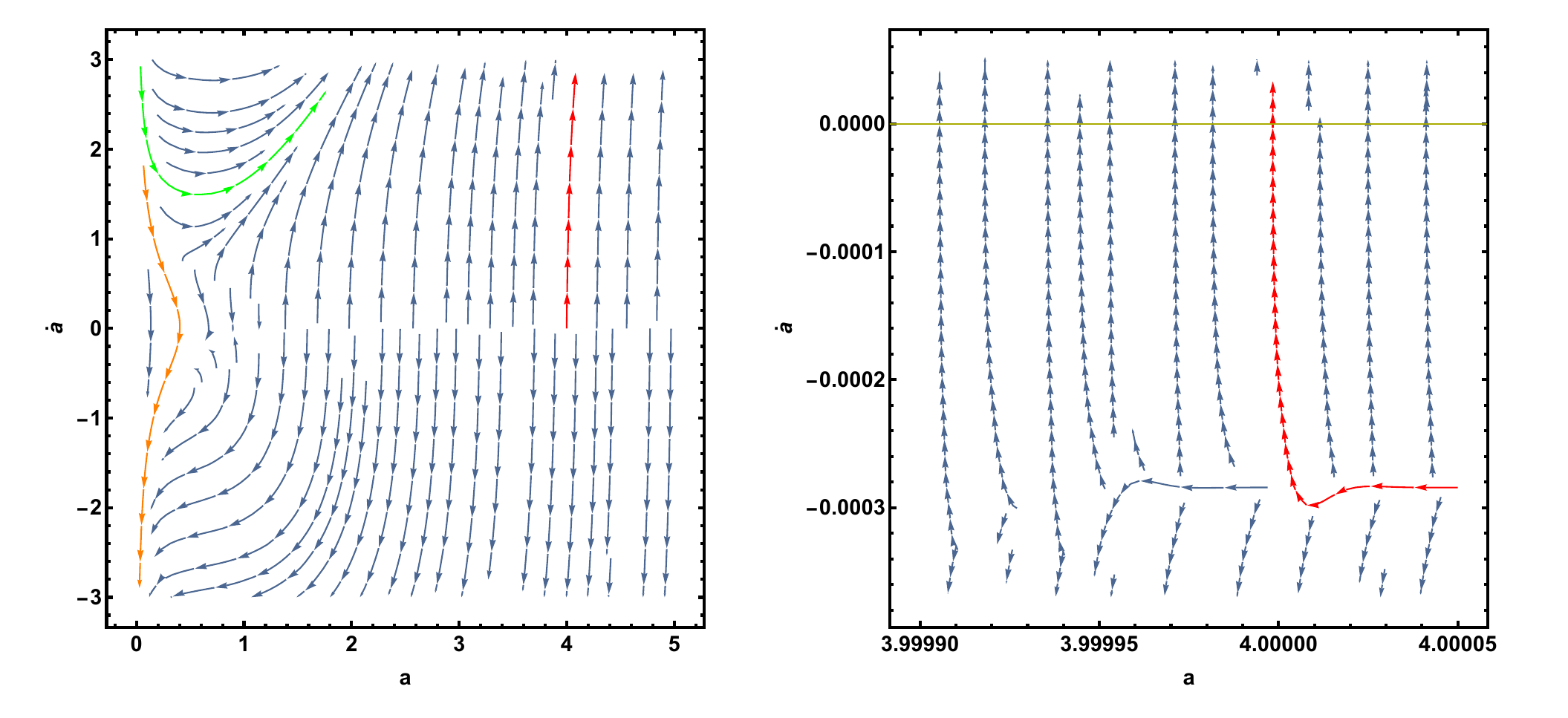}
\caption{Phase diagram of the modified Raychaudhuri equation. The parameter setting is $\kappa=3$, $\Omega_{\rm m}=1/4$, $m_4=m_5=1$, $\mathcal{A}=\mathcal{B}_1=1$, $\Lambda_{\rm eff}=0$ and $K=10^{-3}$, all in units of $\mu=1$. The colored curves are examples of the three classes of solutions that we identify in this model: Big Bang (green), Big Crunch (orange) and non-singular bounce (red). The right panel is a zoomed-in section near the bounce.}
\label{fig:neumann-stream}
\end{figure}

\begin{figure}
	\centering
	\includegraphics[width=0.65\textheight]{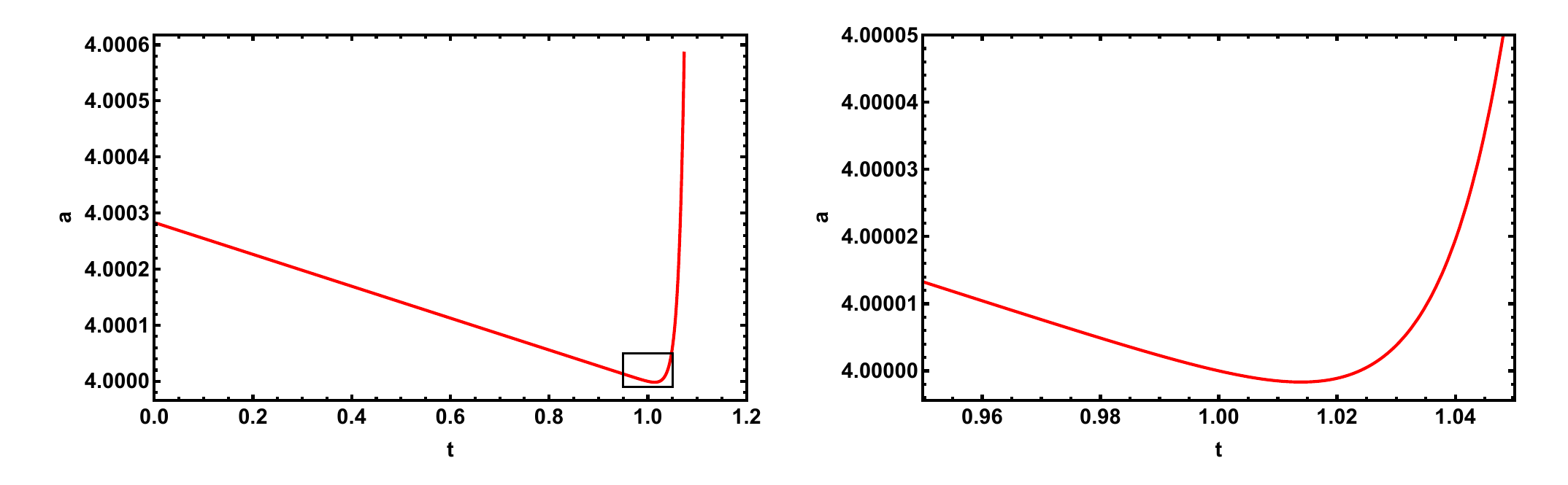}
	\caption{Graph of $a(t)$ for the bounce solution (red curve) of Fig.\ \ref{fig:neumann-stream}. The right panel is a zoomed-in section near the bounce.}
	\label{fig:neumann-bounce}
\end{figure}

Fig.\ \ref{fig:neumann-stream} shows the phase diagram of the modified Raychaudhuri equation for a particular parameter setting. We identify three classes of solutions, illustrated by the particular solutions highlighted in the graph. These include singular Big Bang and Big Crunch type solutions, as well as an interesting and broad class of solutions characterized by a non-singular bounce. The latter however exhibits a Big Rip type singularity, as suggested by the graph of $a(t)$ shown in Fig.\ \ref{fig:neumann-bounce}.

These features of the bounce model may be understood analytically. Eq.\ \eqref{eq:modified raych} becomes trivial in the naive $K\to0$ limit where we neglect all but the term $\propto \dot{a}/\sqrt{K}$. This suggests a different dominant balance in which the second derivative is large, an intuition that is confirmed by inspection of Fig.\ \ref{fig:neumann-stream}. If $\dot{a}$ is also large, then we obtain the approximate equation
\beq \label{eq:bounce1}
\dot{a}\ddot{a}=\lambda a^3\,,\qquad \lambda\equiv \frac{2\mu^2m_5^2\mathcal{B}_1}{\kappa^2\sqrt{K}} \,.
\eeq
This admits an exact solution given by
\beq
a(t)=\frac{36}{\lambda}\left(t_{\rm rip}-t\right)^{-3} \,,
\eeq
which correctly approximates the late-time behavior observed numerically, and in particular confirms the existence of a Big Rip singularity (where we assume $\lambda>0$, i.e.\ $\mathcal{B}_1>0$, as in Fig.\ \ref{fig:neumann-bounce}).\footnote{Alternatively, notice that \eqref{eq:bounce1} may be solved analytically for the Hubble parameter: $H=\left(\frac{3\lambda}{4}\right)^{1/3}\frac{(a^4+c)^{1/3}}{a}$, where $c$ is an integration constant. At late times this leads to $H\propto a^{1/3}$, corresponding to an effective phantom-like fluid with equation of state parameter $w=-11/9$.} If, on the other hand, $\dot{a}$ is assumed small compared to the other scales in the problem, with $a$ kept finite, we get a different approximation given by
\beq \label{eq:bounce2}
a\ddot{a}=K \,.
\eeq
This innocent-looking equation does not admit a closed-form solution in terms of elementary functions. The general solution may be written implicitly in terms of the Dawson function $D_+$ as
\beq \label{eq:dawson bounce}
\pm \sqrt{\frac{K}{2}}(t-t_{\rm b})=aD_+\left(\log^{1/2}\frac{a}{a_{\rm b}}\right) \,,
\eeq
where $t_{\rm b}$ is the time of the bounce and $a_{\rm b}\equiv a(t_{\rm b})$. Indeed, using the fact that $D_+(x)\simeq x$ for small $x$, we find the approximation
\beq
a\simeq a_{\rm b}+\frac{K}{2a_{\rm b}}\left(t-t_{\rm b}\right)^2 \,,
\eeq
valid for $t\simeq t_{\rm b}$, confirming the existence of a bounce. Interestingly, this approximation is completely universal, in the sense that it is independent of all the parameters of the theory but the curvature constant $K$. It is also worth emphasizing that the existence of a class of bounce solutions does not require any exotic matter in this model; in fact, it does not even require matter.

\subsection{Special model and modified Friedmann equation} \label{subsec:special model}

The special model of WMG cosmology is defined by the condition $\mathcal{C}_1=\mathcal{C}_2=0$ on the constants given in \eqref{eq:C coeffs}. This model therefore requires a relation among the coefficients of the 4D and 5D massive gravity potentials. We choose to solve these constraints for $\beta_3$ and $\beta_5$:
\beq\bal \label{eq:special beta sol}
\beta_3&= \alpha_3-\frac{\alpha_4-\beta_4}{1+2\alpha_3-2\sigma\sqrt{1+\alpha_3+\alpha_3^2-\alpha_4}} \,,\\
\beta_5&=-1-\left(1+\alpha_3+\alpha_3^2-\beta_4\right)\left(1+2\alpha_3-2\sigma\sqrt{1+\alpha_3+\alpha_3^2-\alpha_4}\right) \\
&\quad +\alpha_3(\alpha_4-\beta_4)+\alpha_4\left(1+\alpha_3-\frac{\alpha_4-\beta_4}{1+2\alpha_3-2\sigma\sqrt{1+\alpha_3+\alpha_3^2-\alpha_4}}\right) \,.
\eal\eeq
Remarkably, this choice results in $\mathcal{B}_1=0$, i.e.\ the vanishing of the term $\propto m_5^2\frac{\dot{a}}{\sqrt{K}}$ in the modified Raychaudhuri equation. As explained, it is precisely this term which forbids the derivation of a Friedmann-type equation in the Neumann model. With $\mathcal{B}_1=0$ we find that Eq.\ \eqref{eq:modified raych} admits the following first integral:
\beq\bal
\frac{\mathcal{C}}{a^4}&= H^2\left[1+\frac{\kappa^2}{2\mu^2}\left(\frac{K}{a^2}+\frac{\mu\rho-m_4^2\mathcal{A}}{3}\right)\right]-\frac{\kappa^2}{4\mu^2}H^4-\frac{K}{a^2}\left[1+\frac{\kappa^2}{6\mu^2}\left(\mu\rho-m_4^2\mathcal{A}\right)\right] \\
&\quad -\frac{\kappa^2}{36\mu}\rho\left(\mu\rho-2m_4^2\mathcal{A}\right)-\frac{\Lambda_{\rm eff}}{6} \,,
\eal\eeq
and we recall the definition of $\Lambda_{\rm eff}$ in \eqref{eq:lambda eff}. This equation is quartic in $H$, yielding two `Friedmann branches' $H^2(a)$ as in the DGP case.

For simplicity we set $K=0$, finding
\beq \label{eq:modified friedmann}
H^2=\frac{\mu}{3}\rho+2\left(\frac{\mu^2}{\kappa^2}-\frac{m_4^2\mathcal{A}}{6}\right)\pm\frac{2\mu}{\kappa}\sqrt{\frac{\mu}{3}\rho-\frac{\mathcal{C}}{a^4}+\frac{\kappa^2}{\mu^2}\left(\frac{\mu^2}{\kappa^2}-\frac{m_4^2\mathcal{A}}{6}\right)^2-\frac{\Lambda_{\rm eff}}{6}}
\eeq
It is clear from this expression that this model includes an effective cosmological constant, which however differs from the constant term $\Lambda_{\rm eff}$ in the modified Raychaudhuri equation, as previously mentioned. We can make this more manifest by expanding at large $a$ and assuming standard matter on the brane, given for the sake of concreteness by uncoupled dust and radiation, with density parameters $\Omega_{\rm m}$ and $\Omega_{\rm r}$, respectively. This gives
\beq \label{eq:modified friedmann late time}
\frac{H^2}{H_0^2}=\frac{\Omega_{\rm r}'}{a^4}+\frac{\Omega_{\rm m}'}{a^3}+\frac{2}{H_0^2}\left[\left(\frac{\mu^2}{\kappa^2}-\frac{m_4^2\mathcal{A}}{6}\right)\pm\frac{\mu}{\kappa}\sqrt{\frac{\kappa^2}{\mu^2}\left(\frac{\mu^2}{\kappa^2}-\frac{m_4^2\mathcal{A}}{6}\right)^2-\frac{\Lambda_{\rm eff}}{6}}\right]+\mathcal{O}\left(a^{-6}\right) \,,
\eeq
which matches the form of the Friedmann equation of GR, although with `effective' matter density parameters given by
\beq\bal
\Omega_{\rm r}'&=\Omega_{\rm r}\pm\frac{\frac{\mu}{\kappa}\left(\Omega_{\rm r}-\frac{\mathcal{C}}{H_0^2}\right)}{\sqrt{\frac{\kappa^2}{\mu^2}\left(\frac{\mu^2}{\kappa^2}-\frac{m_4^2\mathcal{A}}{6}\right)^2-\frac{\Lambda_{\rm eff}}{6}}} \,,\\
\Omega_{\rm m}'&=\Omega_{\rm m}\pm\frac{\frac{\mu}{\kappa}\,\Omega_{\rm m}}{\sqrt{\frac{\kappa^2}{\mu^2}\left(\frac{\mu^2}{\kappa^2}-\frac{m_4^2\mathcal{A}}{6}\right)^2-\frac{\Lambda_{\rm eff}}{6}}} \,.\\
\eal\eeq
Notice that this assumes $\frac{\kappa^2}{\mu^2}\left(\frac{\mu^2}{\kappa^2}-\frac{m_4^2\mathcal{A}}{6}\right)^2-\frac{\Lambda_{\rm eff}}{6}\neq0$. Otherwise the modified Friedmann equation \eqref{eq:modified friedmann} does not lead to a reinterpretation of the density parameters of matter and radiation, although it still modifies the GR behavior. To illustrate this, let us set $\frac{\mu^2}{\kappa^2}-\frac{m_4^2\mathcal{A}}{6}=0$ and $\Lambda_{\rm eff}=0$, resulting in a late-time behavior for the $+$ branch given by
\beq
\frac{H^2}{H_0^2}=\frac{2\mu\sqrt{\Omega_{\rm m}}}{\kappa H_0}a^{-3/2}+\mathcal{O}\left(a^{-5/2}\right) \,,
\eeq
corresponding to an effective fluid with equation of state parameter $w=-1/2$, which in GR would violate the strong energy condition. If on the other hand we consider a model with pure radiation, we have instead
\beq
\frac{H^2}{H_0^2}=\frac{\Omega_{\rm r}}{a^4}\pm \frac{2\mu\sqrt{\Omega_{\rm r}-\mathcal{C}/H_0^2}}{\kappa H_0}\frac{1}{a^2} \,,
\eeq
i.e.\ an effective curvature term. We remark that these curious models are peculiar to massive gravity and are thus not available in the DGP set-up, which always leads to a dark energy-like behavior at late times, $H^2\propto {\rm constant}$.

\begin{figure}
\centering
\includegraphics[width=0.65\textheight]{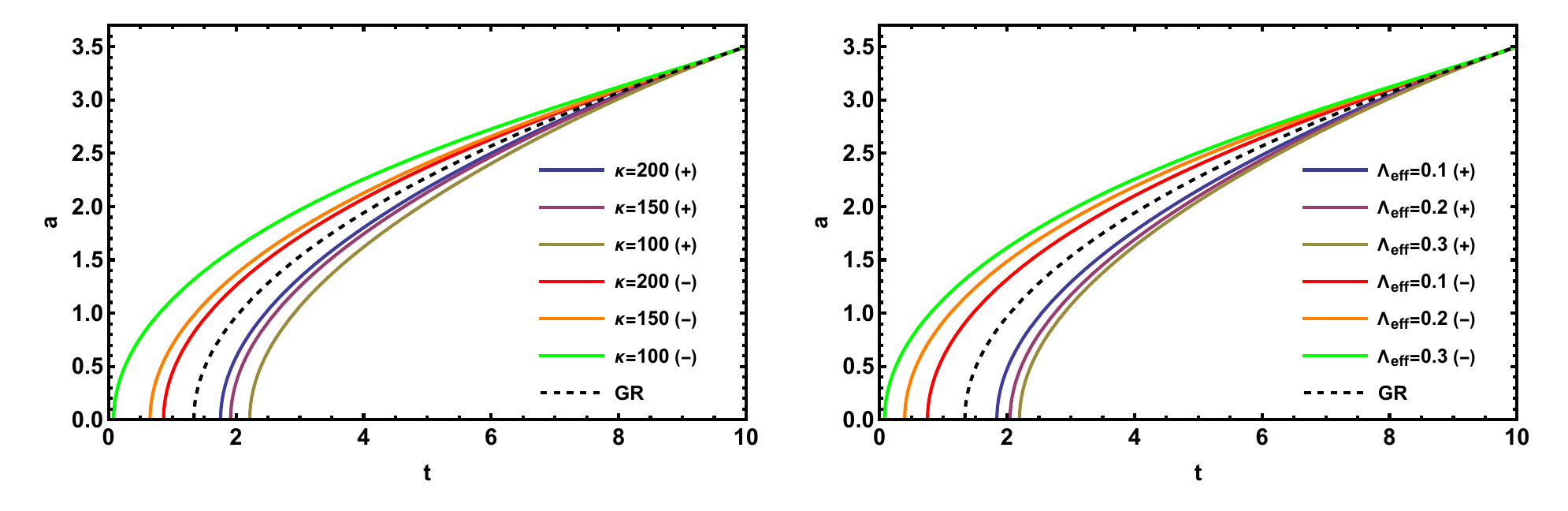}
\caption{Graph of $a(t)$ in the special model for different choices of $\kappa$ (left panel) and $\Lambda_{\rm eff}$ (right). Other parameters are fixed as $m_4=1/10$, $\Omega_{\rm r}=1/2$ and $\mathcal{A}=\pm1/10$ (respectively for the $\pm$ branch), all in units of $\mu=1$. When varying $\kappa$, $\Lambda_{\rm eff}=1/10$; when varying $\Lambda_{\rm eff}$, $\kappa=170$. In each case, $\mathcal{C}$ is chosen such that the slope matches the corresponding GR result at $t=10$.}
\label{fig:special-plots}
\end{figure}

Ignoring such particular cases, it is apparent from \eqref{eq:modified friedmann late time} that WMG only introduces modifications to cosmology at early times in this model, at least at the background level. Fig.\ \ref{fig:special-plots} shows the results of numerically integrating Eq.\ \eqref{eq:modified friedmann} with varying values of $\kappa$ and $\Lambda_{\rm eff}$, for both choices of branches. For simplicity, we focus on a pure-radiation model, and compare the results with the curve of the corresponding GR solution with the same matter content.

%====================================

\section{Observational predictions of the special model} \label{sec:data analysis}

In this section we present a first analysis of the observational predictions and consistency of WMG. As we do not aim here towards an exhaustive treatment, we limit our attention to the special model of Sec.\ \ref{subsec:special model} restricted to a particular parameter setting.

As we have seen, the special model of WMG consistently agrees with the predictions of the $\Lambda$CDM model at late times, at least qualitatively at the background level. Modifications are introduced at early times, which leads us to inquire if WMG has the potential to resolve or at least alleviate the so-called `Hubble tension', namely the significant discrepancy between the value of the Hubble parameter predicted by early-universe data and that measured locally at late times through distance and redshift observations \cite{Knox:2019rjx,Coley:2019yov,DiValentino:2021izs,Abdalla:2022yfr,Cervantes-Cota:2023wet}. Specifically, one calculates the Hubble parameter using the power spectrum of the cosmic microwave background (CMB), a method which however relies on the choice of theoretical model \cite{Kamionkowski:2022pkx}. In contrast, local measurements based on the direct distance-redshift relation typically employ the ``distance ladder'' approach, for instance in the observation of supernovae, yielding a model-independent value of the Hubble parameter. The issue becomes apparent when comparing the results from these two methods, as discrepancies ranging from 4\( \sigma \) to 6\( \sigma \) have been reported \cite{DiValentino:2021izs}.

\subsection{Set-up}

The observational datasets and likelihoods used in this work are as follows:

\begin{itemize}
	\item 
	CMB temperature and polarization anisotropy measurements from the \textit{Planck} satellite. The CMB dataset used in this work
	include:  
	(i) the \textit{Planck}-2018 low multipole ($2 \le \ell\le 30$) temperature anisotropy power spectrum $C^{TT}_{\ell}$, reconstructed using the \texttt{Commander} likelihood; 
	(ii) the \textit{Planck}-2018 low multipole ($2 \le \ell\le 30$) large-scale E-mode polarization power spectrum $C^{EE}_{\ell}$, derived from the \texttt{Simall} likelihood; 
	(iii) high multipole power spectra of temperature and polarization anisotropies, $C^{TT}_{\ell}$, $C^{TE}_{\ell}$, $C^{EE}_{\ell}$, derived from the \texttt{NPIPE PR4 Planck CamSpec} likelihood \cite{Rosenberg:2022sdy}; 
	(iv) the lensing potential power spectrum, derived from the NPIPE PR4 lensing reconstruction data \cite{Carron:2022eyg, Carron:2022eum}. We label the combination of these likelihoods as ``CMB''. 
	\item 
	The PantheonPlusSH0ES supernova catalog, containing PantheonPlus sample \cite{Brout:2022vxf} (contains 1701 light curves recorded from 1550 type Ia supernovae within the redshift domain $0.001 < z < 2.26$) removing the data points in redshift range $z < 0.01$, with anchoring of SN standardized absolute magnitude, SH0ES \cite{Riess:2020fzl}. We will refer to the combination of these supernova data as ``PPS''.
\end{itemize}

In line with our goal of presenting a simple proof of principle, we focus on the modified Friedmann equation \eqref{eq:modified friedmann} with the parameter setting $K=\mathcal{C}=0$. This results in the equation
\begin{equation} \label{eq:data modified friedmann}
\frac{H^{2}}{H_{0}^{2}} =  \frac{\Omega_{\rm r}}{a^4}+\frac{\Omega_{\rm m}}{a^3} +2\left(M^{2}-\frac{m_4^2\mathcal{A}}{6H_0^2}\right)+2M \sqrt{N+ \frac{\Omega_{\rm r}}{a^4}+\frac{\Omega_{\rm m}}{a^3}} \,,
\end{equation}
where 
\beq
M\equiv \pm \frac{{\mu}}{ \kappa H_{0}} \,,\qquad N\equiv \frac{\kappa^2}{\mu^2H_0^2}\left(\frac{\mu^2}{\kappa^2}-\frac{m_4^2\mathcal{A}}{6}\right)^2-\frac{\Lambda_{\rm eff}}{6H_0^2} \,.
\eeq
$\Omega_{\rm m}$ is the present-day matter density parameter, including baryons $\Omega_{\rm b}$, cold dark matter $\Omega_{\rm c}$ and one massive neutrino with $m_{\nu} =0.06 \, {\rm eV}$; $\Omega_{\rm r}$ is the present-day radiation density parameter, including photons and massless neutrinos.

We compute the predictions of our model with the Boltzmann code \texttt{CAMB} \cite{Lewis:1999bs,Howlett:2012mh}.\footnote{In our numerical calculations we use the equation
\beq
\frac{H^{2}}{H_{0}^{2}} -1 =  \frac{\Omega_{\rm r}}{a^4}+\frac{\Omega_{\rm m}}{a^3}-\left(\Omega_{\rm r}+\Omega_{\rm m}\right) +2M\left[ \sqrt{N+ \frac{\Omega_{\rm r}}{a^4}+\frac{\Omega_{\rm m}}{a^3}} -\sqrt{N+ \Omega_{\rm r}+\Omega_{\rm m}}\right] \,,
\eeq
obtained from \eqref{eq:data modified friedmann} by replacing $2\left(M^{2}-\frac{m_4^2\mathcal{A}}{6H_0^2}\right) = 1- \Omega_{\rm r}-\Omega_{\rm m} -
2M \sqrt{N+ \Omega_{\rm r}+\Omega_{\rm m}}$, which follows from evaluating at the present time $t_0$ (with the normalization $a(t_0)\equiv 1$).} As for the $\Lambda$CDM model, the standard six-parameter basis is used: the physical densities of baryons ($\Omega_{\rm b} h^2$) and cold dark matter ($\Omega_{\rm c} h^2$), the Hubble constant ($H_0$) inferred from the approximation to the acoustic angular scale ($\theta_{\rm MC}$), the optical depth ($\tau_{\rm reio}$), the amplitude of the primordial scalar power spectrum ($\ln ( 10^{10} A_{\rm s} )$) and its spectral index ($n_{\rm s}$). A summary of cosmological parameters and their priors for different models are provided in Table \ref{table: prior}. We consider four `models' of WMG defined by different priors.

\begin{table}[t]
\centering
\begin{tabular}{cccccc}
	\toprule
	\hline
	\multicolumn{2}{c}{Model} &Parameter&
	\multicolumn{3}{c}
	{Prior}\\
	\hline
	\multirow{6}{*}
	{$\Lambda$CDM}&
	\multirow{6}{*}
	& $\Omega_{\rm b} h^2$                            & norm  & $0.0222$  & $0.0005$  \\
	&& $\Omega_{\rm c} h^2$                            & flat  & $0.001$   & $0.99$    \\
	&& $100\,\theta_{\rm MC}$                     & flat  & $0.5$     & $10$   \\
	&& $\ln ( 10^{10} A_{\rm s} )$                     & flat  & $1.61$    & $3.91$    \\
	&& $n_{\rm s}$                                     & flat  & $0.8$     & $1.2$     \\
	&& $\tau_{\rm reio}$                        & flat  & $0.01$    & $0.8$     \\
	\hline
	\multirow{5}{*}
	{\text{WMG}}
	&
	& $N$   & flat  & $0$  & $1$ \\
	&M& $M$   & flat  & $-1$  & $1$       \\
	&$\text{M(+)}$& $M$   & flat  & $0$  & $1$    \\
	&$\text{M}(-)$& $M$   & flat  & $-1$  & $0$    \\
	&$\text{M}(-0.05)$& $M$   & flat  & $-1$  & $-0.05$ \\
	\hline
	\bottomrule
\end{tabular}
\caption{The model parameters of $\Lambda$CDM and four WMG models, distinguished by their priors on the parameter $M$. For the normal prior distribution (first row) the two last columns indicate the mean and the width, respectively. For the flat distributions the two last columns indicate the minimum and maximum.}
\label{table: prior}
\end{table} 

We utilize the \texttt{Cobaya} Markov Chain Monte Carlo (MCMC) sampler to generate the posterior distribution of the full set of cosmological parameters \cite{Torrado:2020dgo, Lewis:2002ah, Lewis:2013hha, 2005math......2099N}. For each dataset, we consider the chains as converged when the Gelman-Rubin stopping criterion $R-1 < 0.01$ is satisfied. We use \texttt{GetDist} \cite{Lewis:2019xzd} to carry out a statistical evaluation of MCMC samples and plot the posteriors in our investigation. The baryon density $\Omega_{\rm b} h^2$ is sampled from the BBN prior and a cut $20 < H_0 < 100$ is imposed on the Hubble constant, as inferred from the angular size of the sound horizon, $\theta_{\rm MC}$ \cite{Hu:1995en}.

We emphasize that linear perturbations are treated as in $\Lambda$CDM, without any modification. This point is irrelevant as far as the PPS dataset is concerned, but it is of course not self-consistent for the CMB data analysis.
We regard such a background-only approach to be appropriate for a first analysis and considering that a complete study of cosmological perturbations in WMG is outside the scope of the present paper. We may expect it to provide a reasonable approximation, at least in some region of the parameter space, or serve as a theoretically motivated phenomenological model, in a similar spirit to what is typically done with bottom-up parametrizations of dynamical dark energy for the purpose of model comparison; see e.g.~\cite{DESI:2025zgx}.

\subsection{Results}

To evaluate differences between datasets and models, we utilize the traditional Bayesian evidence (B), the deviance information criterion (DIC) and the Watanabe-Akaike information criterion (WAIC) for model comparison. We introduce 
$\Delta -\ln\text{B} \equiv -\ln\text{B}_{\rm model}-(-\ln\text{B}_{\Lambda \rm CDM}) $, $\Delta \text{DIC} \equiv \text{DIC}_{\rm model}-\text{DIC}_{\Lambda \rm CDM} $ and $\Delta \text{WAIC} \equiv \text{WAIC}_{\rm model}-\text{WAIC}_{\Lambda \rm CDM} $ to compare each model with $\Lambda$CDM.

Our results of the analysis are presented in Table \ref{Results} and Appendix \ref{app:data plots}. We observe that there is a tendency in the data to favor the $\Lambda$CDM model. However, for the four WMG models we tested, the relative information criteria and Bayes factor do not exceed 1.62. The results therefore fall within a credible range.\footnote{We remark that these results should be interpreted with caution, as our posterior distributions are quite irregular, cf.\ Appendix \ref{app:data plots}. In this regard, the WAIC should be seen as the most robust estimator among the three, and the fact that our results for $\Delta$DIC and $\Delta$WAIC are fairly consistent suggests a meaningful model comparison.} Remarkably, for models with negative values of $M$, the comparison of the Hubble constant calculated from CMB and late-time data reveals that the discrepancy between the two methods is significantly smaller than in $\Lambda$CDM, and is in fact fully resolved within $1\sigma$ in the M$(-0.05)$ model where $M$ is assumed strictly negative as a prior. On the other hand, from the Table and Appendix \ref{app:data plots}, we observe that $M$ and $N$ remain rather poorly constrained. In particular, they are both consistent with their $\Lambda$CDM values for the models we considered. Moreover, we note that the alleviation of the Hubble tension appears to come at the price of introducing a significant discrepancy in the measurements of $\Omega_{\rm m}$ inferred from early- and late-universe data. This is a manifestation of the well-known degeneracy between $H_0$ and $\Omega_{\rm m}$ \cite{Zaldarriaga:1997ch,Efstathiou:1998xx} (see also \cite{Colgain:2022nlb,Colgain:2022rxy}).

\begin{table}
	\centering
	\resizebox{\columnwidth}{!}{%
	\def\arraystretch{1.25}
	\begin{tabular}{ccccccccc}
		\toprule
		\hline
		Dataset & Model  & $H_{0}$ & $\Omega_{\rm m}$ & $M$  & $N$ &$\Delta-\ln\text{B}$&$\Delta \text{DIC}$ &$\Delta \text{WAIC}$ \\
		\midrule
		\multirow{5}{*}{CMB} &
		{$\Lambda$CDM}  & $67.21\pm 0.46$ & $0.3158\pm 0.0063$ & $0$&$0$  &$0$&$0$&$0$\\&
		{$\text{M}$}  & $58\pm 8$ & $0.455^{+0.078}_{-0.17}$ & $ 0.29^{+0.20}_{-0.32}$& $ --$ &$0.55$&$0.54$& $0.81$ \\&
		{$\text{M} (+)$}  & $56^{+9}_{-6}$ & $0.473^{+0.059}_{-0.15}$ & $  < 0.426$& $--$ &$-2.03$&$0.14$&$0.05$\\&
		{$\text{M} (-)$}  & $70.7^{+1.3}_{-3.5}$ & $0.287^{+0.028}_{-0.014}$ & $  > -0.0950$& $--$ &$-1.75$&$0.77$&$0.57$\\&
		{$\text{M}(-0.05)$}  & $72.3^{+1.0}_{-3.0} $ & $0.273^{+0.023}_{-0.011}$ & $> -0.126$& $--$ &$-1.13$&$1.44$&$1.54$\\
		\midrule
		\multirow{5}{*}{PPS} & $\Lambda$CDM  & $73.5\pm 1.0$ & $0.332\pm 0.018$ & $0$& $0$ &$0$&$0$&$0$\\
		&$\text{M}$  & $73.6\pm 1.0$ & $0.298^{+0.059}_{-0.13}$ & $0.22\pm 0.40$& $> 0.388$ &$1.62$&$0.46$&$0.77$\\
		&$\text{M} (+)$ & $73.5\pm 1.0$ & $0.238^{+0.048}_{-0.068}$ & $< 0.599$& $--$  &$0.53$&$0.02$&$0.04$\\
		&$\text{M} (-)$ & $73.7\pm 1.0$ & $0.416^{+0.038}_{-0.083}$ & $ > -0.285$& $> 0.418$  &$1.31$&$1.13$&$1.06$\\
		&$\text{M} (-0.05)$ & $73.7\pm 1.0$ & $0.429^{+0.034}_{-0.086}$ & $> -0.309$& $> 0.438$  &$0.97$&$1.13$&$1.03$\\
		\hline
		\bottomrule
	\end{tabular}
}
\caption{Cosmological parameter constraints for different data combinations and priors, in the $\Lambda$CDM and the four WMG models studied in this work. Results are presented for the marginalized mean and $68\%$ confidence intervals in each case. $H_0$ is in units of $\mathrm{km} \, \mathrm{s}^{-1}\,\mathrm{Mpc}^{-1}$. The blanks indicate that a meaningful constraint is not available due to the flatness of the resulting probability distribution, cf.\ Appendix \ref{app:data plots}.}
\label{Results}
\end{table}

%====================================

\section{Discussion} \label{sec:conclusion}

We have performed a first study of the cosmological predictions of the theory of WMG. From a 4D perspective, WMG corresponds to a class of models rather than a single one, since different solutions in the bulk can in principle lead to qualitatively distinct descriptions of the cosmology on the brane. Contrary to the DGP model, in WMG there is no unique way to obtain independent equations for the brane dynamics, as these depend on the boundary conditions for the St\"uckelberg fields. In turn, the equations of motion for the St\"uckelberg fields, i.e.\ the massive gravity constraints, cannot in general be decoupled from the bulk dynamics as they involve the function $b$. Nevertheless, we have shown that particular choices of boundary conditions and/or parameters lead to self-consistent models of 4D cosmology.

The Neumann model of WMG appears to be distinctly novel in that the resulting evolution equation for the scale factor depends non-perturbatively on the spatial curvature scale. Moreover, we have argued that this set-up does not admit a `standard' modified Friedmann equation, again in stark contrast with the DGP model. Our numerical and analytical analysis has revealed an intriguing type of solutions characterized by a smooth bounce yet without the need of exotic matter. On the other hand, the special model of WMG is qualitatively similar to DGP, although it does allow for novel types of behaviors for particular choices of parameters. The confrontation with observational data shows that this set-up provides a correct fit with credibility at a comparable level as the standard $\Lambda$CDM model. Interestingly, the theory is capable of resolving the tension between early- and late-time measurements of the Hubble constant, although at the price of generating a tension in the observed matter density.

It is worth summarizing our scope and main assumptions, which may be suggest avenues for follow-up research:
\begin{itemize}
	\item Following the original WMG proposal, we assumed the 4D induced fiducial metric to be flat. In turn, this led us to the familiar obstacle for constructing flat FLRW solutions in massive gravity. Here we followed the simplest resolution, i.e.\ to consider open FLRW metrics. It should be emphasized, however, that this no-go result for flat FLRW cosmology relies on an assumption on the spacetime foliations defined by the physical and fiducial metrics, and it may be a priori possible that more general parametrizations could evade the obstacle. Alternatively, the extension of the WMG set-up to bigravity \cite{Hassan:2011zd} is an intriguing possibility.
	
	\item The field equations on the brane depend on the 5D metric function $b$, on the St\"uckelberg functions $f$ and $g$, in addition to the scale factor $a$. Although $b$ may be eliminated upon imposing the junction conditions, similarly to the DGP set-up, one is still left with the jumps $[f']$ and $[g']$. We have assumed the simplest choice of boundary condition here, namely $[f']=0$, which nicely also cancels any dependence on $[g']$. On the other hand, once this choice is made, the constraint equations imply that either $[g']=0$ (Neumann model) or else the need of a special relation among the parameters of the 5D and 4D massive gravity potentials (special model).
	
	\item As our primary interest rests on the 4D cosmology, we have not attempted to solve the bulk equations of motion. Related to this, recall that we assumed a $\mathbb{Z}_2$-symmetry about the brane, which is equivalent to regarding the brane as a boundary of the 5D space. More generally, one could consider asymmetric boundary conditions \cite{Koyama:2005br,Gabadadze:2006jm}.
	
	\item Our analysis was restricted to the background cosmological level. The study of perturbations is certainly important, in particular to determine the conditions of stability. This question is especially pertinent given the known instabilities that arise in the context of 4D massive gravity~\cite{DeFelice:2012mx}. We also did not attempt to address whether the solutions of the Neumann model are physically viable.
	
	\item Similarly, for the confrontation with data we made use of a restricted version of the special model. At least at the background level, this analysis could straightforwardly be extended to include all parameters. Once again, however, a full and self-consistent assessment of the viability of the theory would necessarily require to take into account cosmological perturbations.
\end{itemize}

\subsubsection*{Acknowledgments}

We are very grateful to C\'edric Deffayet for collaboration, guidance and helpful conversations. We also thank Fotios Anagnostopoulos and Huang Ling for comments on the draft. The authors acknowledge support from the NSFC (Grant No.\ 12250410250). SGS also acknowledges support from a Provincial Grant (Grant No.\ 2023QN10X389).

\appendix

\section{First integral in the Neumann model} \label{app:first integral}

We have claimed that the modified Raychaudhuri equation \eqref{eq:modified raych} does not admit a ``simple'' first integral, by which we mean that this integral cannot be a simple enough function of $H$ and $a$ which would allow one to solve for $H$ in terms of $a$ in closed form, thus furnishing a Friedmann-type equation. In this appendix we provide a proof of this based on a simplified form of the equation. This is sufficient as our claim applies to generic choices of parameters, while we cannot rule out the possibility that particular, but non-trivial, parameter settings may produce an integrable equation.\footnote{Indeed, notice that the two limiting cases investigated in Sec.\ \ref{subsec:neumann model}, Eqs.\ \eqref{eq:bounce1} and \eqref{eq:bounce2}, do possess simple first integrals.}

We consider the equation
\beq
\frac{\ddot{a}}{a}+H^2-\frac{K}{a^2}+\sigma \dot{a}=0 \,,
\eeq
where $\sigma$ is a constant, corresponding to the (artificial) limit where we set the jump functions and effective cosmological constant to zero in \eqref{eq:modified raych}. This may be rewritten as
\beq \label{eq:app1}
\frac{dH}{da}=-2\frac{H}{a}+\frac{K}{Ha^3}-\sigma \,.
\eeq
When $\sigma=0$, this equation has the first integral $\mathcal{C}=H^2a^4-Ka^2$. We seek the correction to this in the limit of small $\sigma$, i.e.\ we solve \eqref{eq:app1} perturbatively to first order in $\sigma$. We find
\beq
\mathcal{C}=H^2a^4-Ka^2+\frac{\sigma a^4}{4K^{3/2}}\left[\sqrt{K}Ha\left(H^2a^2+K\right)+\left(H^2a^2-K\right)^2\log\left|a\big(Ha-\sqrt{K}\big)\right|\right] +\mathcal{O}(\sigma^2) \,.
\eeq
It is clear that, truncating at this order, we obtain an expression that cannot be solved for $H(a)$ in closed form. Although we cannot discard the possibility that a resummation of this expansion may be possible, this seems unlikely. In fact, Eq.\ \eqref{eq:app1} may be recast in the form of a Chini equation, for which exact solutions are only available in special cases, and \eqref{eq:app1} does not belong to this class.\footnote{The class of known solutions of the general Chini equation is characterized by a constant value of the so-called Chini invariant \cite{Kamke2011}. One can check that the Chini invariant is not constant for Eq.\ \eqref{eq:app1}.} This excludes the possibility of analytically integrating the modified Raychaudhuri equation for generic values of parameters.

\section{Constraints on cosmological parameters} \label{app:data plots}

In this Appendix we collect the plots for the constraints on cosmological parameters resulting from the data analysis of Sec.\ \ref{sec:data analysis}, limited to the parameters $H_0$, $\Omega_{\rm m}$, $M$ and $N$ (cf.\ \eqref{eq:data modified friedmann}).
\begin{figure}
	\centering
	\includegraphics[width=0.5\textheight]{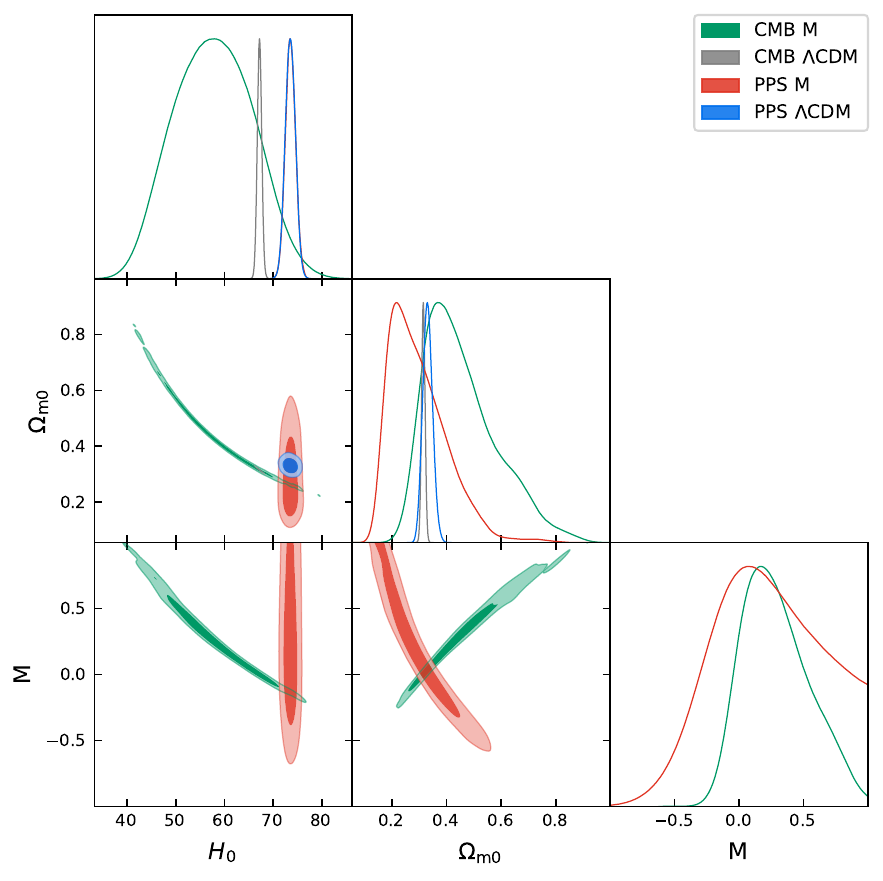}
	\caption{Constraints on parameters for $\Lambda$CDM and the WMG special ``M'' model assuming a flat prior $M\in[-1,1]$. Shown are $1\sigma$ and $2\sigma$ contours obtained from the CMB and PPS datasets.}
\label{fig:dataM}
\end{figure}
\begin{figure}
	\centering
	\includegraphics[width=0.5\textheight]{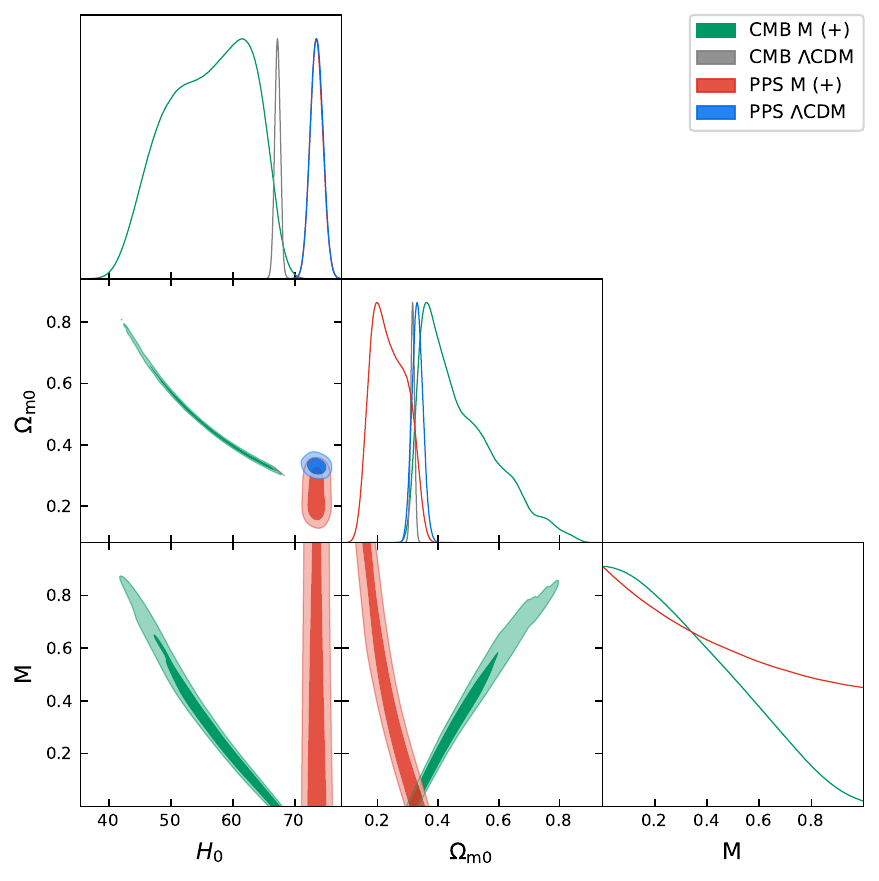}
	\caption{Same as Fig.\ \ref{fig:dataM} but assuming a flat prior $M\in[0,1]$ for the special ``M'' model.}
%\label{fig:data+}
\end{figure}
\begin{figure}
	\centering
	\includegraphics[width=0.5\textheight]{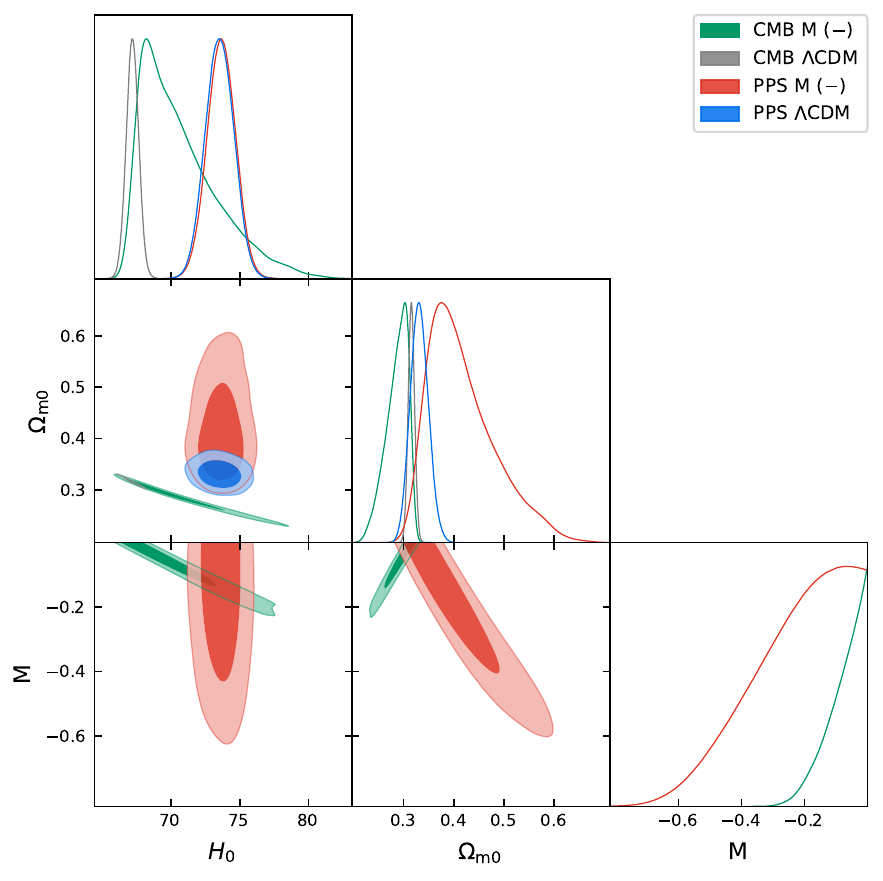}
	\caption{Same as Fig.\ \ref{fig:dataM} but assuming a flat prior $M\in[-1,0]$ for the special ``M'' model.}
%\label{fig:data-}
\end{figure}
\begin{figure}
	\centering
	\includegraphics[width=0.5\textheight]{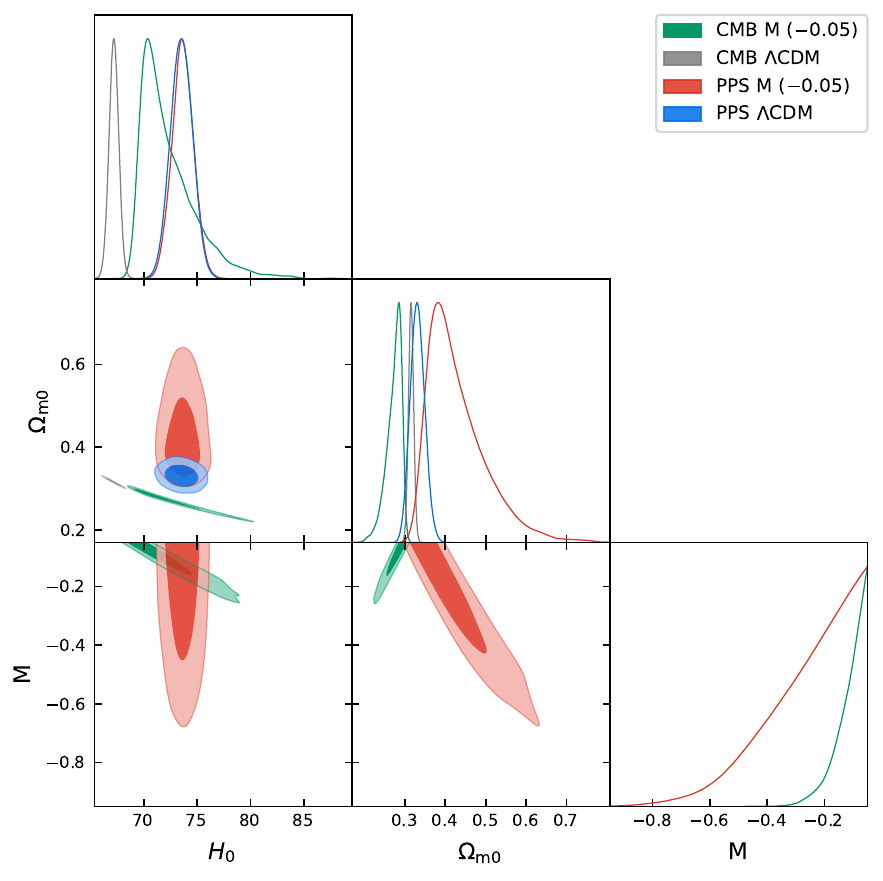}
	\caption{Same as Fig.\ \ref{fig:dataM} but assuming a flat prior $M\in[-1,-0.05]$ for the special ``M'' model.}
%\label{fig:data-005}
\end{figure}
\begin{figure}
	\centering
	\includegraphics[width=0.5\textheight]{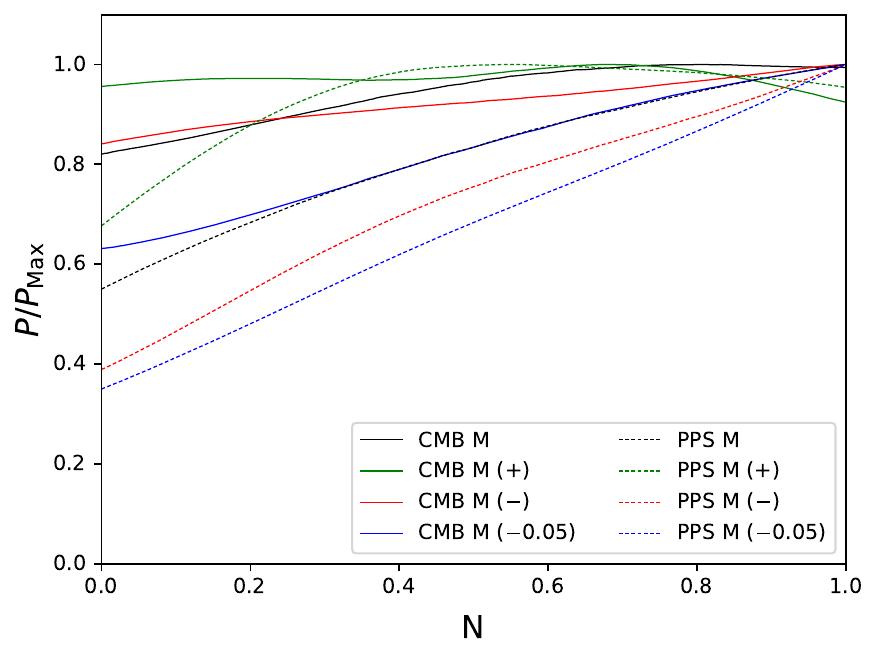}
	\caption{Constraints on the parameter $N$ obtained from the CMB and PPS datasets, for the four special ``M'' models studied in this work.}
	%\label{fig:}
\end{figure}

%====================================

\bibliographystyle{unsrturl}
\bibliography{WMGcosmology_refs}

\end{document}